# The family of quantum droplets keeps expanding

Addition of lattice potentials helps to produce new species of stable fundamental and vortical quantum droplets in two dimensions

Boris A. Malomed[1,2]

[1]Department of Physical Electronics, School of Electrical Engineering, Faculty of Engineering, and Center for Light-Matter Interaction, Tel Aviv University, P.O.B. 39040, Tel Aviv, Israel

[2]Instituto de Alta Investigación, Universidad de Tarapacá, Casilla 7D, Arica, Chile

In the course of the past 25 years, several new quantum states of matter have been created in ultracold gases, starting from the celebrated Bose-Einstein condensates (BECs), which were first made in 1995, following the theoretical prediction published in 1924 [1]. This achievement was followed by the creation and detailed exploration of degenerate Fermi gases [2], Tonks-Girardeau gas of hard-core bosons [3], and spin-orbit-coupled BEC [4]. A recent addition to this set of fundamental states of quantum matter is the prediction [5,6] and experimental demonstration [7,8] of *quantum droplets* (QDs), built of coherent atomic waves in binary (two-component) BEC. This is an extension of BEC beyond the limits of the usual mean-field (MF, alias semi-classical) approximation [1], with the averaged action of quantum fluctuations around the MF states (known as the *Lee-Huang-Yang effect*) leading to drastic changes in static and dynamical properties of the quantum gas. The result may be summarized as an effective quartic nonlinear term, $|\psi|^3\psi$ in Eq. (1), competing with the usual cubic MF nonlinearity in the fundamental Gross-Pitaevskii equation for the macroscopic complex wave function [5], $\psi$, which is written here in the scaled form:

$$i\frac{\partial \psi}{\partial t} = -\frac{1}{2}\nabla^2\psi - g|\psi|^2\psi + |\psi|^3\psi, \qquad (1)$$

where $g > 0$ is the coefficient of the MF interaction. A remarkable advantage of QDs is that the beyond-MF quartic self-repulsion makes it possible to stabilize self-trapped three-dimensional (3D) and quasi-2D droplets, which are kept together by the usual cubic MF attraction between two BEC components [5,6,7], or by the dipole-dipole attraction of atoms carrying magnetic moments [8], against the collapse (catastrophic self-compression of the condensate [9]). In the absence of the beyond-MF repulsion, 3D and quasi-2D matter-wave solitons are inevitably destroyed by the collapse in the framework of the MF approximation.

In the limit when the BEC is strongly confined by an external potential in one direction, Eq. (1) is replaced by the effective 2D equation [6] (see also Ref. [10])

$$i\frac{\partial \psi}{\partial t} = -\frac{1}{2}\nabla^2\psi + \ln\left(|\psi|^2\right)|\psi|^2\psi, \qquad (2)$$

which also supports QDs. It is important to identify conserved quantities (dynamical invariants) supported by Eq. (2). These are the total norm, $N = \iint |\psi(x,y)|^2\,dxdy$, Hamiltonian (energy),

$$H = \frac{1}{2}\iint\left[|\nabla\psi(x,y)|^2 + |\psi|^4 \ln\left(\frac{|\psi|^2}{\sqrt{e}}\right)\right]dxdy, \qquad (3)$$

the linear momentum, and the angular momentum,

$$M = \frac{i}{2}\iint \psi^*\left(x\frac{\partial\psi}{\partial y} - y\frac{\partial\psi}{\partial x}\right)dxdy. \qquad (4)$$

Note that at values of the density $|\psi|^2 < 1$ the nonlinear term in Eq. (2) is self-attractive, which drives the formation of 2D solitons. On the other hand, the change of the sign of $\ln(|\psi|^2)$ at $|\psi|^2 > 1$ switches the self-trapping nonlinear term into the self-repulsive one. Thus, the transition to collapse, that implies unchecked growth of the density, is arrested by the structure of the nonlinearity in Eq. (2), which secures the arrest of collapsing and, eventually, provides the possibility of the creation of *stable* 2D solitons as solutions to Eq. (2).

It is relevant to mention that the reduction of the effective dimension, by means of the transverse tightly confining potential, from 3D to 1D keeps the usual MF cubic term in the resulting amended Gross-Pitaevskii equation, adding to it a quadratic beyond-MF term, $\sim|\psi|\psi$, with the sign which, on the contrary to the 3D and 2D equations (1) and (2), has a self-attractive sign [6], thus it may help to create the respective quasi-one-dimensional QDs [11].

The theoretical and experimental results which demonstrate the existence of stable 3D and quasi-2D QDs [5-8] provide fundamentally important contributions to the vast area of studies of multidimensional solitons, where very few experimental results were available prior to the advent of QDs [12]. Further, it was recently predicted that stable 3D [13] and quasi-2D [14-19] QDs can be created with embedded vorticity. In this connection, it is relevant to stress that prediction and creation of stable 3D and 2D vortex solitons is well known to be an especially challenging problem, for the theoretical and experimental studies alike [20]. As a relevant illustration, Fig. 1 displays examples of stable and unstable profiles of the local density, $|\psi(x,y)|^2$, in stable vortex solitons with winding numbers (topological charges) $S = 1, 2, 3$ (stable) and $S = 4$ (unstable). Note that the angular momentum (4) of the soliton is determined by $S$ and the norm, $M = SN$.

A very recent brief review of the recent experimental and theoretical results for QDs, in both 3D and effectively 2D settings, is given in Ref. [21]. The review addresses the condensates with contact and dipole-dipole interactions. The theoretical part includes results for QDs with embedded vorticity, which have not yet been created in the experiment.

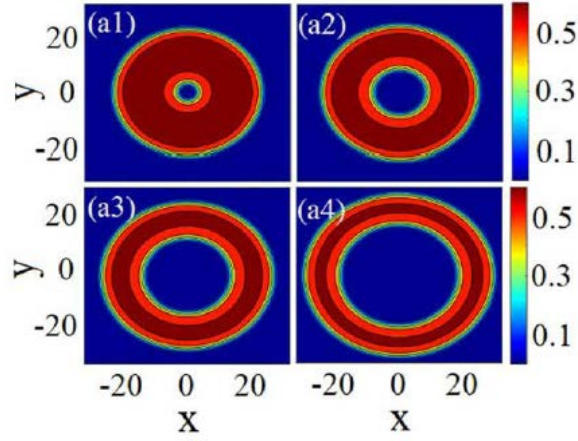

**Fig. 1**. Density profiles of vortex solitons generated by Eq. (2), in polar coordinates $(r,\theta)$, in the form of $\psi = \exp(-i\mu t + iS\theta)\phi(r)$ with chemical potential $\mu$ and integer winding number $S = 1, 2, 3, 4$, shown as per Ref. [15].

Another ingredient of various 2D and 3D models which helps to stabilize zero-vorticity and vortical solitons is a spatially periodic (lattice) potential [22,23]. In the experiment, such a potential can be readily induced in the form of an *optical lattice*, i.e., a spatially periodic force exerted onto atoms in BEC by a resonant optical field, created by pairs of laser beams illuminating the condensate in opposite directions [1,24]. A paper, just now published in Frontiers of Physics [25], makes an important step forward in the theoretical analysis of 2D QDs, by adding a lattice potential, with spatial periodicities $D$ in the $x$ and $y$ directions, to Eq. (2). The potential is represented by the term $V_0\left[\cos^2(\pi x/D) + \cos^2(\pi y/D)\right]\psi$. Systematic numerical analysis of the model has produced new families of soliton solutions, with $S = 0$ and $S = 1$. The solitons, i.e., localized (self-trapped) states, are effectively characterized by the number of excited cells of the underlying lattice, i.e., cells in which the local density is essentially different from zero.

It is relevant to mention that 1D models which combine the above-mentioned self-attractive beyond-MF term and a 1D lattice potential (that may also be induced in the form of an optical lattice) were recently introduced in Refs. [26] and [27]. Those models, in particular, make it possible to predict the existence of stable dipole modes, i.e., bound states of fundamental 1D solitons with opposite signs [26], as well as stable multipeak modes, which may be considered as bound states of a large number of fundamental solitons [27]. The multipeak modes bifurcate from delocalized Bloch states of the underlying linearized system with the periodic potential [26]. These modes with different numbers of peaks realize *multistability*, in the sense that they may coexist as stable solutions for the same value of the chemical potential. On the other hand, the dipole modes may be considered as 1D counterparts of 2D vortex solitons [26]. The mobility, i.e., a possibility of robust motion of a kicked 1D soliton across the underlying lattice potential, was also demonstrated in Ref. [26]. A full quantum many-body treatment of the 1D model with the lattice potential, including the prediction of QDs in this case, was recently elaborated in Ref. [28].

Note that, unlike 2D axisymmetric solutions in free space, such as ones displayed in Fig. 1, the presence of the lattice potential does not make it possible to find axisymmetric solutions, and the potential, breaking the spatial uniformity and isotropy, also breaks the conservation of the linear and angular momenta. Nevertheless, the winding number $S$ of vortex solitons, found as solutions of Eq. (2) including the lattice potential, can be defined as phase circulation of the stationary complex solution, produced by a circumferential trip around the vortex' pivot, divided by $2\pi$ [22,23].

Solitons of both types, with $S = 0$ and $1$, have been constructed in two varieties, *viz.*, onsite and intersite-centered (OC and IC) ones. The OC modes have their pivot placed at a particular site of the underlying lattice, while the IC states place their central point between sites. Stability of the solitons of all these types was identified by means of direct simulations of Eq. (2) for the evolution of perturbed solitons (stable solitons keep their shape, while unstable ones are destroyed by perturbations). Thus, stability areas for the solitons are delineated in the parameter space of the model. Further, the consideration of values of Hamiltonian (3) of the modes demonstrates that the zero-vorticity states are non-degenerate ones, in terms of the energy, while the states with $S=1$ are degenerate, in the sense that two different vortex modes with equal numbers of excited sites may have equal energies.

In addition to that, the lattice potential makes it possible to construct more general localized solutions, which may be interpreted as bound states of fundamental solitons. Such complexes may feature stable asymmetric shapes, in comparison with the symmetry of the underlying lattice potential.

The work may be developed by considering mobility of the solitons, which is a nontrivial issue in the presence of the trapping lattice potential, which tends to suppress the mobility, by means of the corresponding Peierls-Nabarro potential [29]. A challenging direction for the extension of the work can be the consideration of the full 3D model, by adding a spatially periodic potential to Eq. (1), and constructing various solutions of that equation. On the other hand, it may be relevant too to consider the 2D model with an axisymmetric potential, instead of its spatially periodic lattice counterpart. Such a system conserves angular momentum (4), making it relevant to construct axisymmetric solutions for vortex solitons, and may also make it possible to find compact solitons off-shifted from the center and performing circular motion along a circular (or elliptic) trajectory, cf. Ref. [30].